\RequirePackage{fix-cm}
\documentclass[twocolumn]{svjour3}          
\smartqed  
\usepackage{graphicx}
\usepackage{natbib}
\usepackage{color}
\usepackage{marvosym}
\usepackage{ulem}
\begin{document}

\title{Experimental study of a three dimensional cylinder-filament system}




\author{Nicolas Brosse \and Carl Finmo  \and Fredrik Lundell \and Shervin Bagheri
}


\institute{N. Brosse (\Letter), C. Finmo, F. Lundell, S. Bagheri \at
              Linn\'e FLOW Centre, KTH Mechanics, \\ Royal Institute of
							Technology, 100 44 Stockholm, Sweden \\
              \email{brosse@mech.kth.se}           
           }

\date{Received: date / Accepted: date}

\maketitle

\begin{abstract}

This experimental study reports on the behavior of a filament attached to the rear of a three-dimensional cylinder. The axis of the cylinder is placed normal to a uniform incoming flow and the filament is free to move in the cylinder wake. The mean position of the filament is studied as a function of the filament length $L$. It is found that for long ($L/D > 6.5$, where $D$ is the cylinder diameter) and short ($L/D < 2$) filaments the mean position of the filament tends to align with the incoming flow, whereas for intermediate filament lengths ($2 < L/D < 6.5$) the filament lies down on the cylinder and tends to align with the cylinder axis.
The underlying mechanism of the bifurcations are discussed and related to buckling and inverted-pendulum-like instabilities.

\keywords{Wake \and Symmetry breaking \and Fluid-Structure interaction}
\end{abstract}

\section{Introduction}

Many animals, plants and seeds have appendages or protrusions that through a passive interaction with the surrounding fluid can aid locomotion. 
The underlying physical mechanisms responsible for such positive fluid-structure interaction can be isolated using simple body-appendage configurations.
For example,  two-dimensional numerical  \citep{Xual1990} and experimental \citep{CimbalaChen1994} studies have reported on a cylinder free to rotate around its axis with a rigid splitter plate attached to it. When this system is placed in a uniform incoming flow, it is found that for long lengths of the splitter plate, the mean position of the splitter plate is symmetric - i.e. it aligns with the incoming flow - whereas for short lengths, the symmetry is broken, i.e. the mean position of the splitter plate will form an angle with the incoming flow.
More recently, \cite{Bagheri2012} studied through two-dimensional numerical simulations the behavior of a single elastic filament hinged on the rear of a fixed bluff body and placed in a uniform incoming flow. 
They also found a symmetry-breaking  for sufficiently short filament lengths and reported on the induced lift/side force acting on the cylinder-filament system. More specifically, they showed that  for a long filament (symmetric position), the mean lift force and torque acting on the cylinder is zero, whereas when the filament length is reduced below two cylinder diameters, a significant mean lift force/torque acting on the cylinder is generated. 
Even more recently, \cite{Lacis2014}  performed numerical simulations of a free-falling circular cylinder with a splitter plate and found that for short splitter plates the body drifts -- as a consequence of symmetry breaking --  either to the left or right.  They  also proposed a theoretical model - referred to as the inverted-pendulum-like (IPL) model - that provides quantitative predictions of the critical filament length of the symmetry breaking. In this model, the pressure in the recirculation zone behind the bluff body acts as a destabilizing force and is responsible for the symmetry breaking. 
\citet{Lacis2014} also found numerically the IPL instability in a three-dimensional configuration consisting of a sphere (free to rotate and drift) with an attached thin elliptic-shaped sheet. It was confirmed that the mean position of the short sheet forms an angle with the incoming flow and thus drifts. In summary, previous studies have shown that flexible/rigid appendages can  be used to passively control the flow, for example by modifying the trajectory of falling bodies such as cylinders \citep{Horowitz_Williamson_2010}, spheres or disks \citep{Ern_al_2012}.
The aim of the present study is to experimentally investigate the behavior of a three-dimensional cylinder with a filament attached to its rear and placed in a uniform flow. To the best  knowledge of the authors, this configuration has not been investigated previously, as earlier work has either been two-dimensional or numerical. 
The physical problem depends on seven dimensional parameters: the cylinder diameter $D$, the length of the filament $L$, the fluid density $\rho_f$ and filament density $\rho_s$, the bending stiffness of the filament $B$, the free-stream velocity $U$ and the kinematic viscosity $\nu$. Four non-dimensional parameters may then be introduced, namely, the length ratio $L_s=L/D$, the density ratio $\rho=\rho_s/\rho_f$, the Reynolds number $Re = U D / \nu$ and the bending ratio $R_1= B/(\rho_f U^2 D^4)$. In the present study, two free-stream velocities ($U=0.92 \, \rm{cm.s}^{-1}$  and $U=1.56 \, \rm{cm.s}^{-1}$)  are investigated, which results in two configurations ($Re=185$, $R_1 \approx 7.4 \times 10^{-4}$) and ($Re=310$, $R_1 \approx 2.6 \times 10^{-4}$). For each of these configurations,  the length ratio is varied  $0.5\leq L_s\leq 15$, but the density ratio is kept fixed ($\rho=1.25$).

\section{Method}

\begin{figure}
	\begin{center}
		\includegraphics[width=0.4\textwidth]{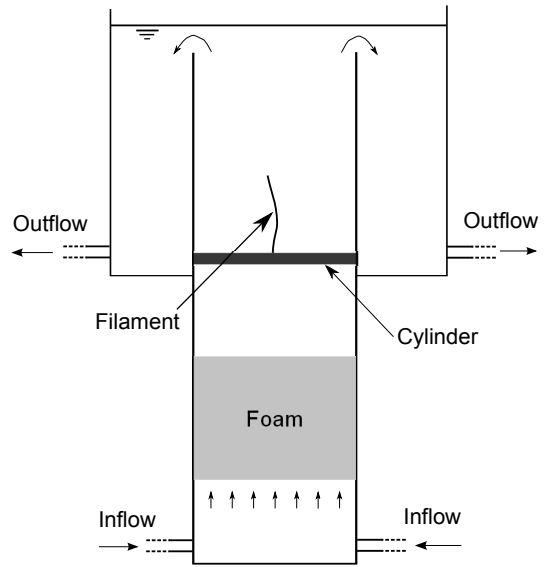}
	\end{center}
\caption{Schematic of the vertical water tunnel facility.}
\label{fig:SchematicExperiment} 
\end{figure}

The experiments have been carried out in a vertical water tunnel; a schematic of the experiment is presented in figure \ref{fig:SchematicExperiment}. 
 The whole apparatus is about $1.5$ m
high, 1 m wide and 0.4 m deep. The dimensions of the test section are 20.5 cm wide,
39.7 cm deep, 68 cm high. Below the test section a 40 cm long slab of
foam is placed in order to homogenize the flow. The  vertical flow of water is generated with a pump with typical flow rate of 0.76 L.s$^{-1}$,  which corresponds to an upward fluid velocity $U = 1$ cm.s$^{-1}$ in the test section.
The cylinder is made from a plexiglass tube with a  diameter  $D = 20$ mm and length of $20.5$ cm. This cylinder is placed in the test section with its axis of symmetry normal to the incoming flow.
 A hole is drilled in the cylinder through which a thread made from silk is inserted. 
The flow inside the channel was previously measured by \cite{Bellani2005} with ultrasonic velocity profiler (UVP).  At the position of the cylinder -- 20 cm above the foam -- fluid velocity profiles for $U_z$ were measured along the $y$ direction; the fluid velocity $U_z$ was found to vary  within  $\pm5\%$ over $50\%$ of the center of the channel, the turbulence intensity of $U_z$  was found to be $5\%$.

\begin{figure}
	\begin{center}
		\includegraphics[width=0.2\textwidth]{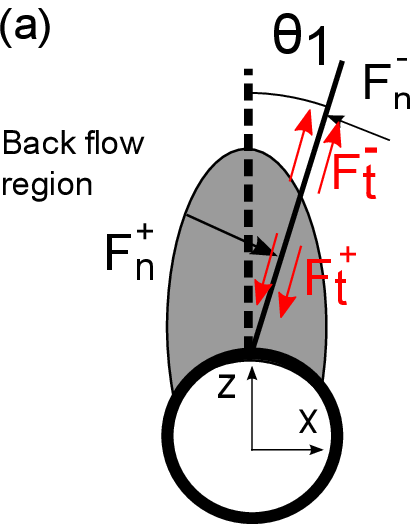} \hspace{8mm}
		\includegraphics[width=0.2\textwidth]{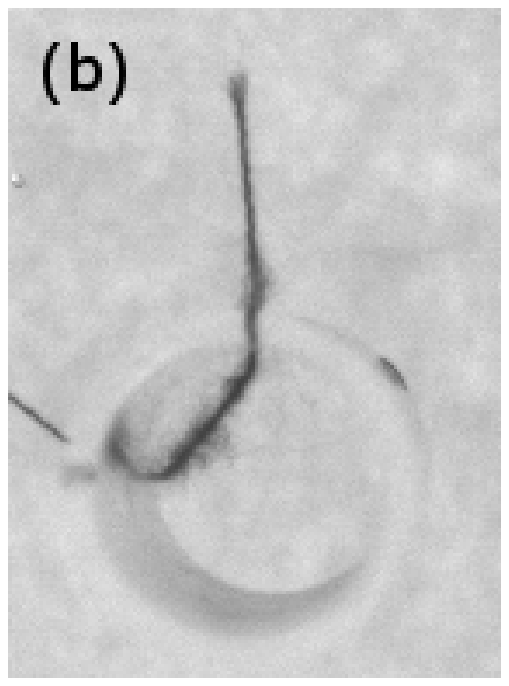}
	\end{center}
\caption{ (a) Schematic of the cylinder-filament for the front
view The grey region marks the back-flow region. Inside this region the flow is in reverse direction to the free stream. The normal forces on the filament inside $F_n^+$ and outside $F_n^-$ the back-flow region and the tangential forces $F_t^+$ and $F_t^-$ are also shown; (b) photography of cylinder-filament system for the front view.}
\label{fig:Schematic_Angles_Theta_1}
\end{figure}

\begin{figure}
	\begin{center}
		\includegraphics[width=0.3\textwidth]{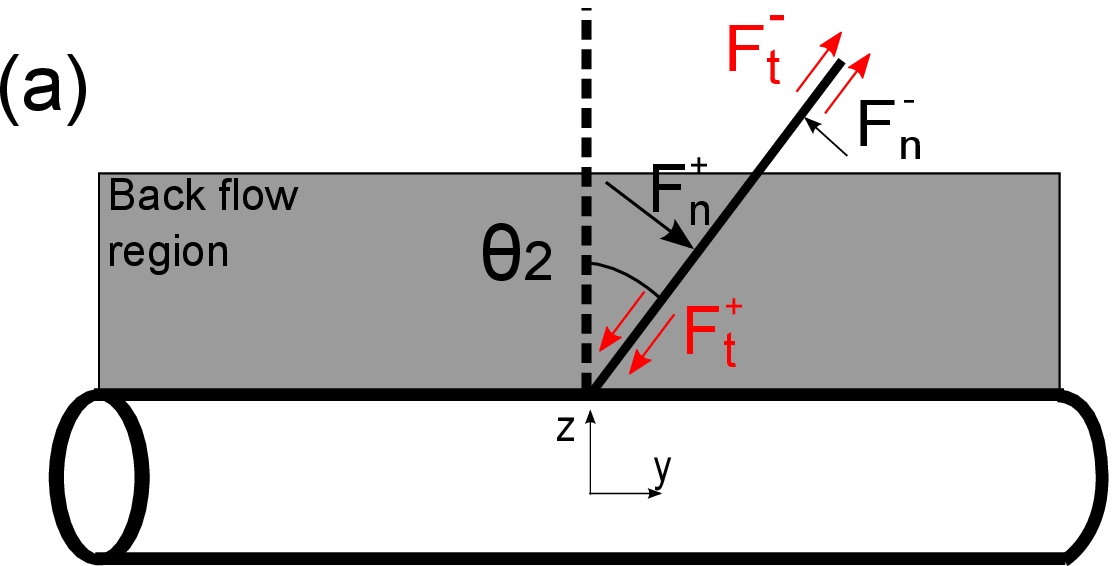}\\
	 \vspace{4mm}
		\includegraphics[width=0.3\textwidth]{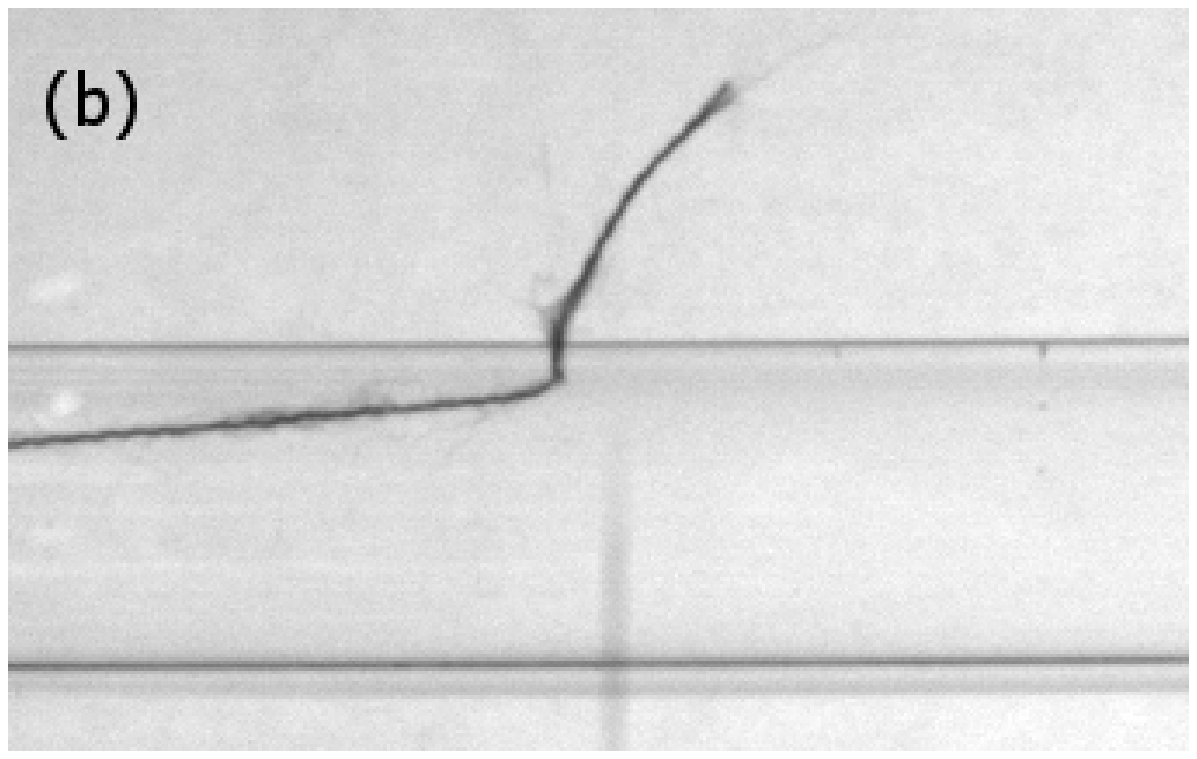}
	\end{center}
\caption{ (a) Schematic of the cylinder-filament for the side view; (b)  photography of cylinder-filament system for the side view.}
\label{fig:Schematic_Angles_Theta_2}
\end{figure}

The position of the filament can be defined by two angles of deflection,
one from the front and one from the side - called $\theta_1$ and $\theta_2$, respectively. These
two angles are shown schematically in figures \ref{fig:Schematic_Angles_Theta_1}(a) and \ref{fig:Schematic_Angles_Theta_2}(a).
As the shape of the filament is not a straight line but can be curved it can be noted that the two angles of deflection $\theta_1$ and $\theta_2$ do not take into account the shape of the filament. As a consequence, the two angles give only partial information regarding the position of the filament, nevertheless $\theta_1$ and $\theta_2$ indicate the behavior of the filament regarding the symmetry breaking and the lie down behavior that this paper focuses on.
In order to measure the position of the filament, the cylinder is photographed
with two perpendicular cameras ($1400\times1088$
pixels), one from the front and one from the side. The conversion factor is found to be 5 pixels per mm corresponding to an accuracy of $\pm 0.1$ mm in the determination of the filament position. 
For each length of the filament a series of grey scale images are recorded; in total seventeen filament lengths are recorded between $0.5 \leq L_s \leq 15$. For each series an acquiring rate of 2 Hz and an exposure time of 10 ms  was used; for the Reynolds number $Re=185$ the recording time is 4 minutes and 10 seconds resulting in 500
images and for $Re=310 $ the recording time is 8 minutes and 20 seconds resulting in 1000
images for each measured filament length.
The greyscale images are then converted to black and white images in order to detect the filament position.
The angles of deflection are calculated as the angles between the vertical and the line formed by the
filament attachment point to the cylinder and the endpoint of the filament (positive angles are defined in the clockwise direction).
To obtain the mean angles of deflection, the angles for the entire
measurement series (500 and 1000 images for $Re=185$ and $Re=310$ respectively) are averaged in time and the absolute value of this mean is then taken to give the mean angles of deflection $\langle\theta_1\rangle$ and $\langle\theta_2\rangle$.

\section{Filament behavior}

\begin{figure}
	\begin{center}
		\includegraphics[width=0.5\textwidth]{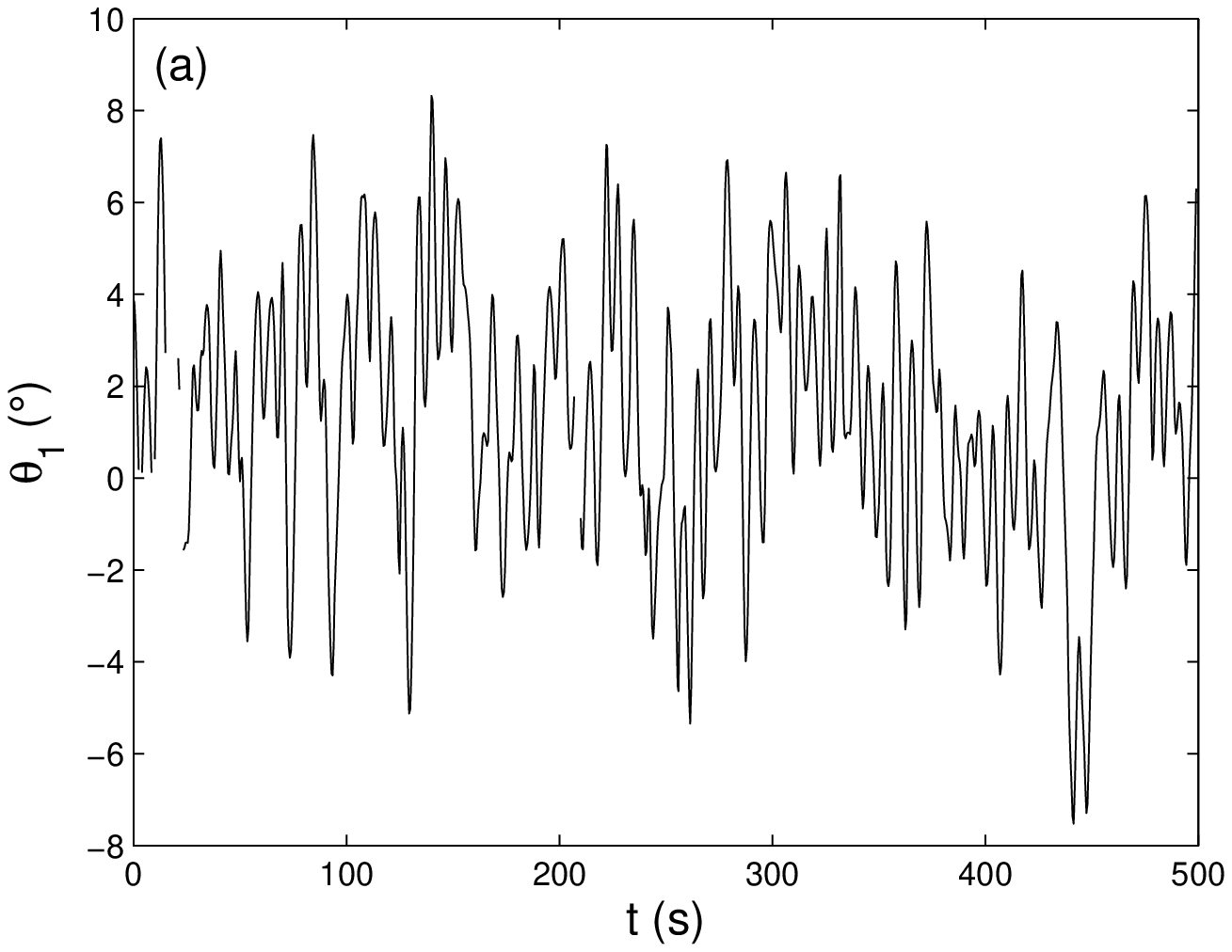}
		\includegraphics[width=0.5\textwidth]{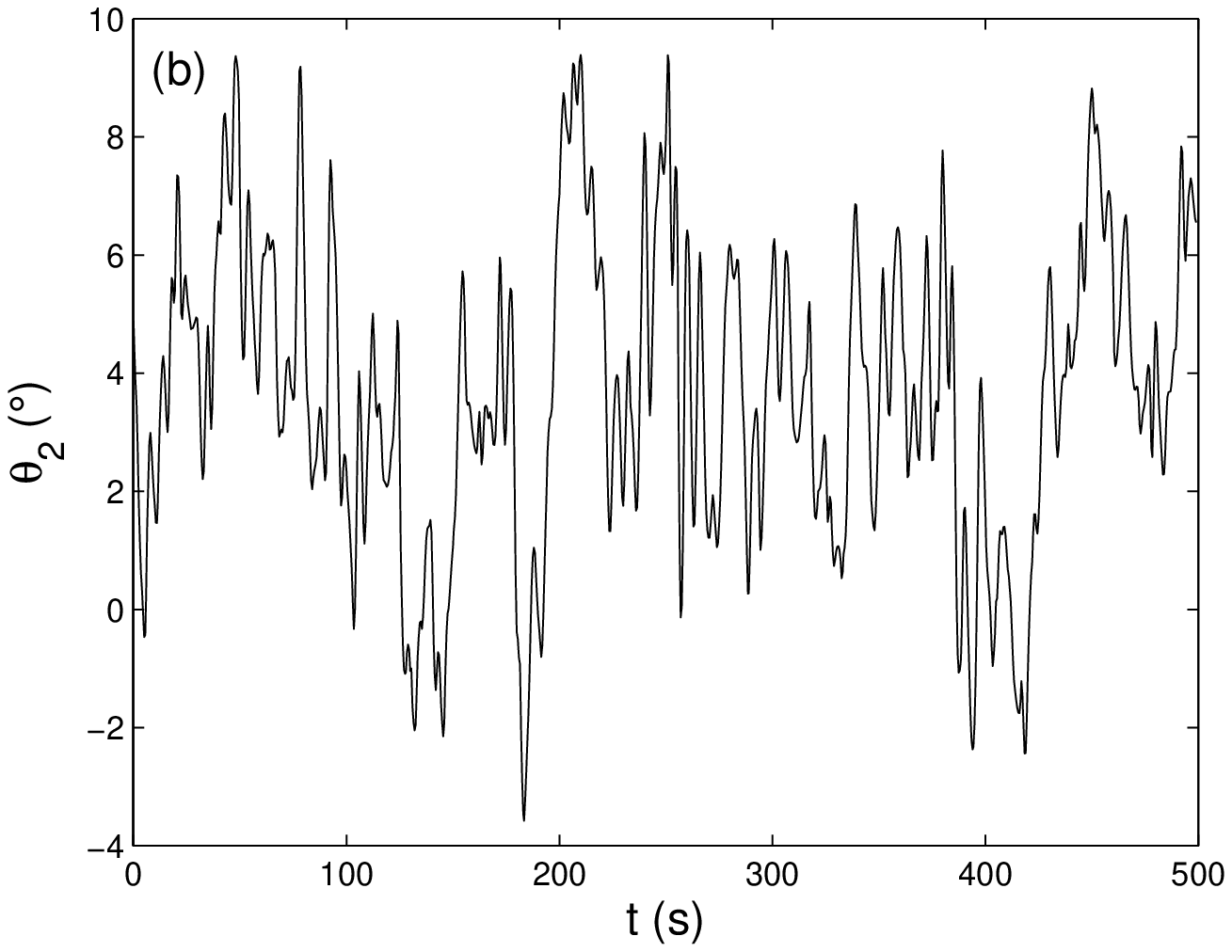}
		\includegraphics[height=0.45\textwidth]{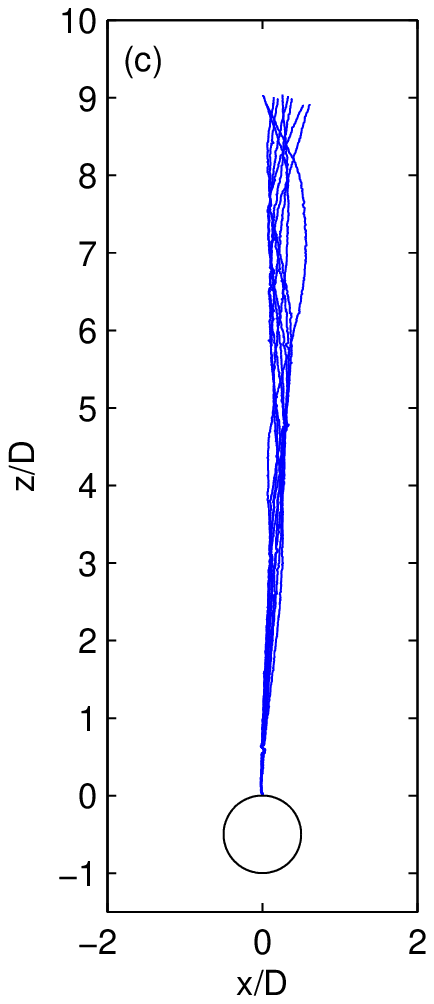}
		\includegraphics[height=0.45\textwidth]{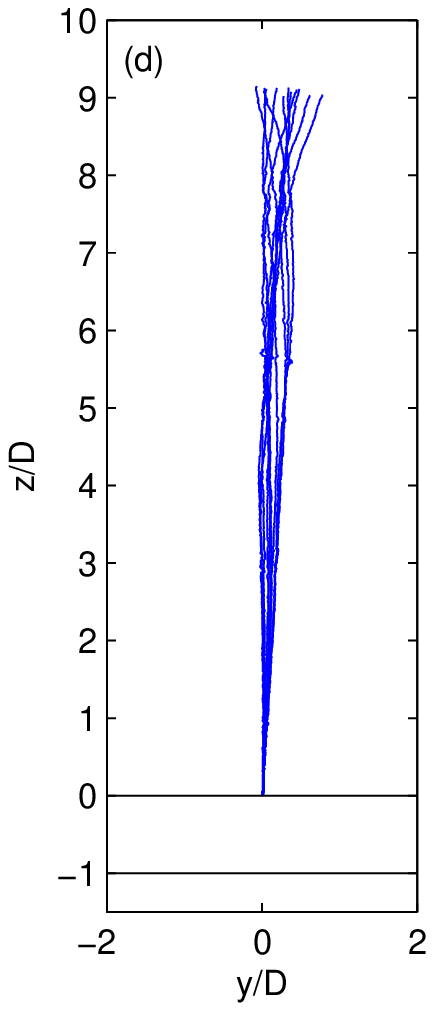}
		\end{center}
\caption{The angles of deflection for (a) $\theta_1$ and (b) $\theta_2$ as a function of time and filament position at different times ($t=0$ s to $t=10$ s) for (c) the front view and (d) the side view; $L/D = 9.2$ and $Re=310$.}
\label{fig:Theta_vs_t_L9.2_Re310}
\end{figure}

\begin{figure}
	\begin{center}
		\includegraphics[width=0.5\textwidth]{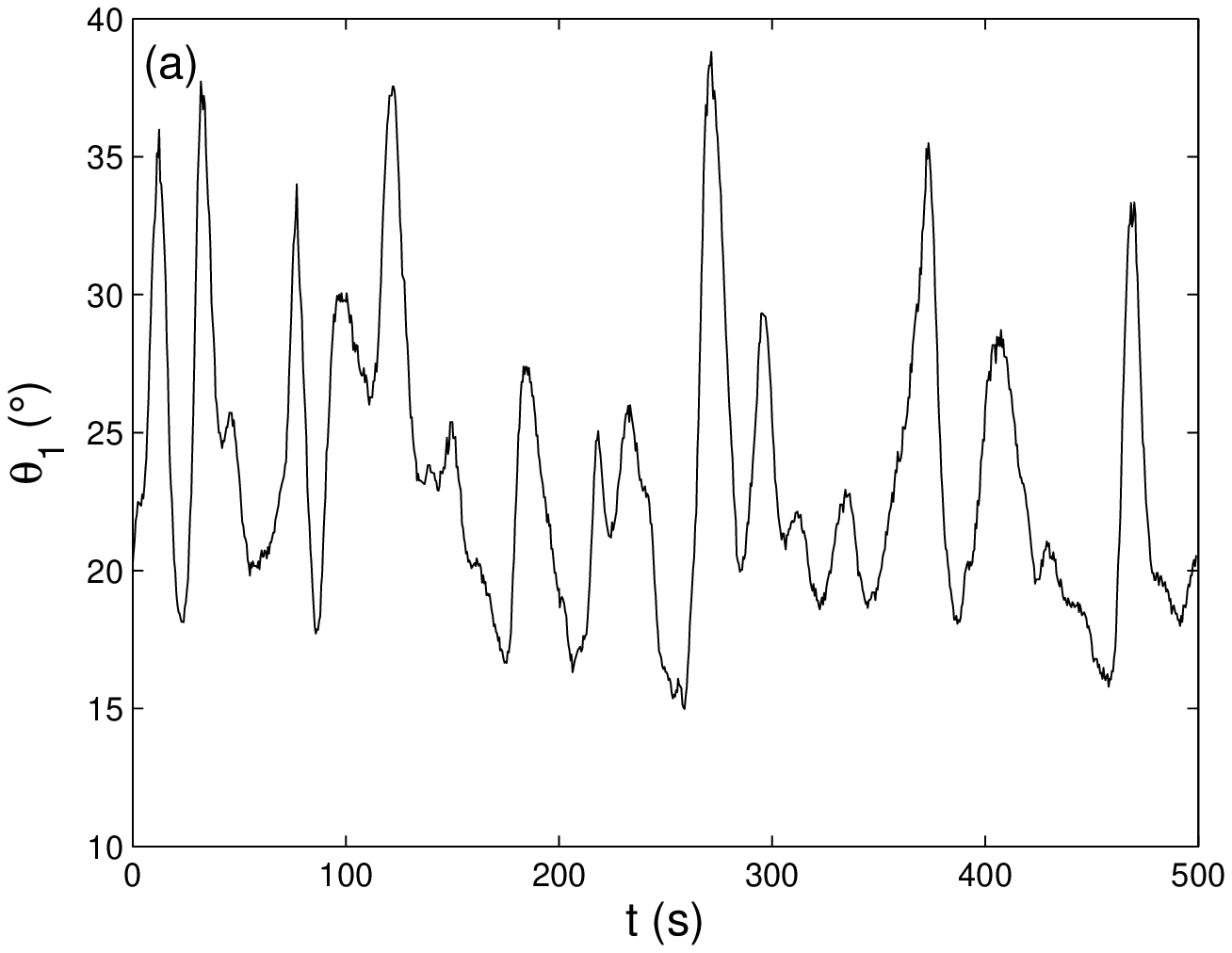}
		\includegraphics[width=0.5\textwidth]{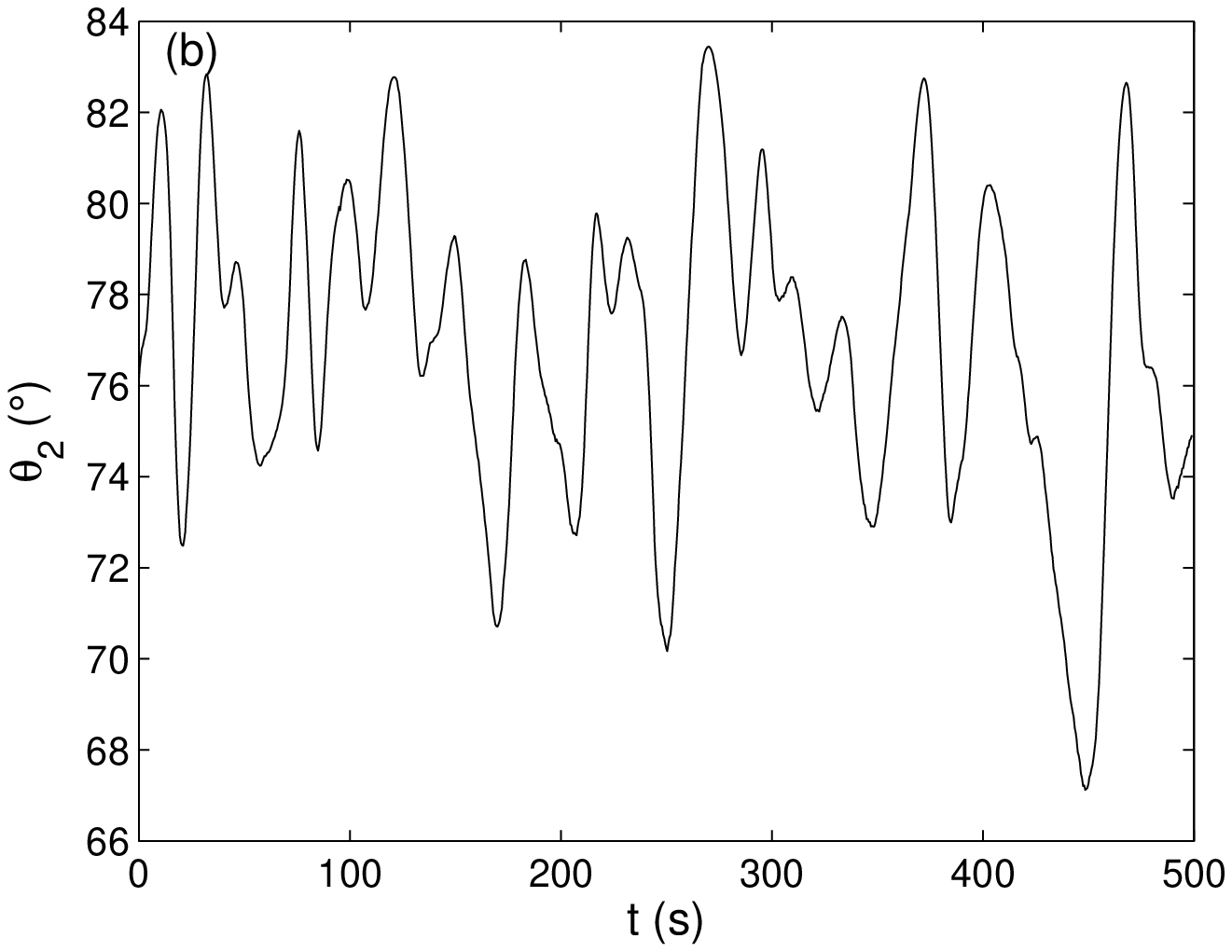}
		\includegraphics[height=0.18\textwidth]{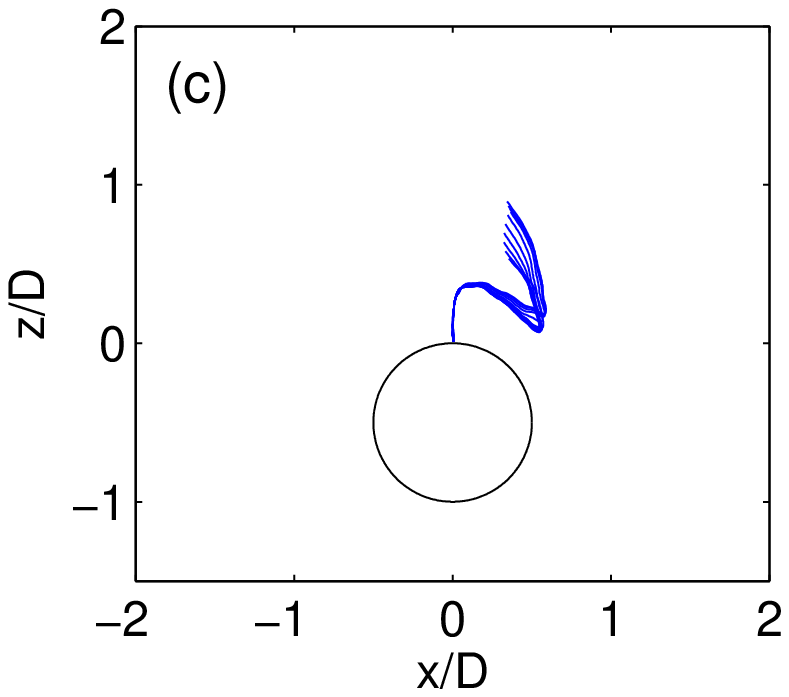}
		\includegraphics[height=0.18\textwidth]{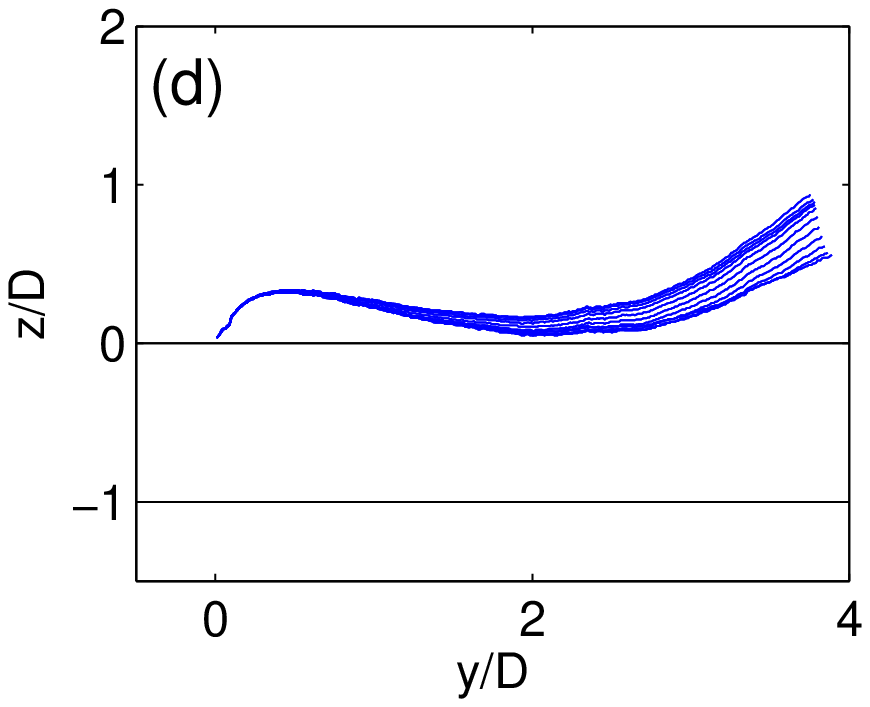}
		\end{center}
\caption{Same as figure \ref{fig:Theta_vs_t_L9.2_Re310} but for $L/D = 4.1$.}
\label{fig:Theta_vs_t_L4.08_Re310}
\end{figure}

\begin{figure}
	\begin{center}
		\includegraphics[width=0.5\textwidth]{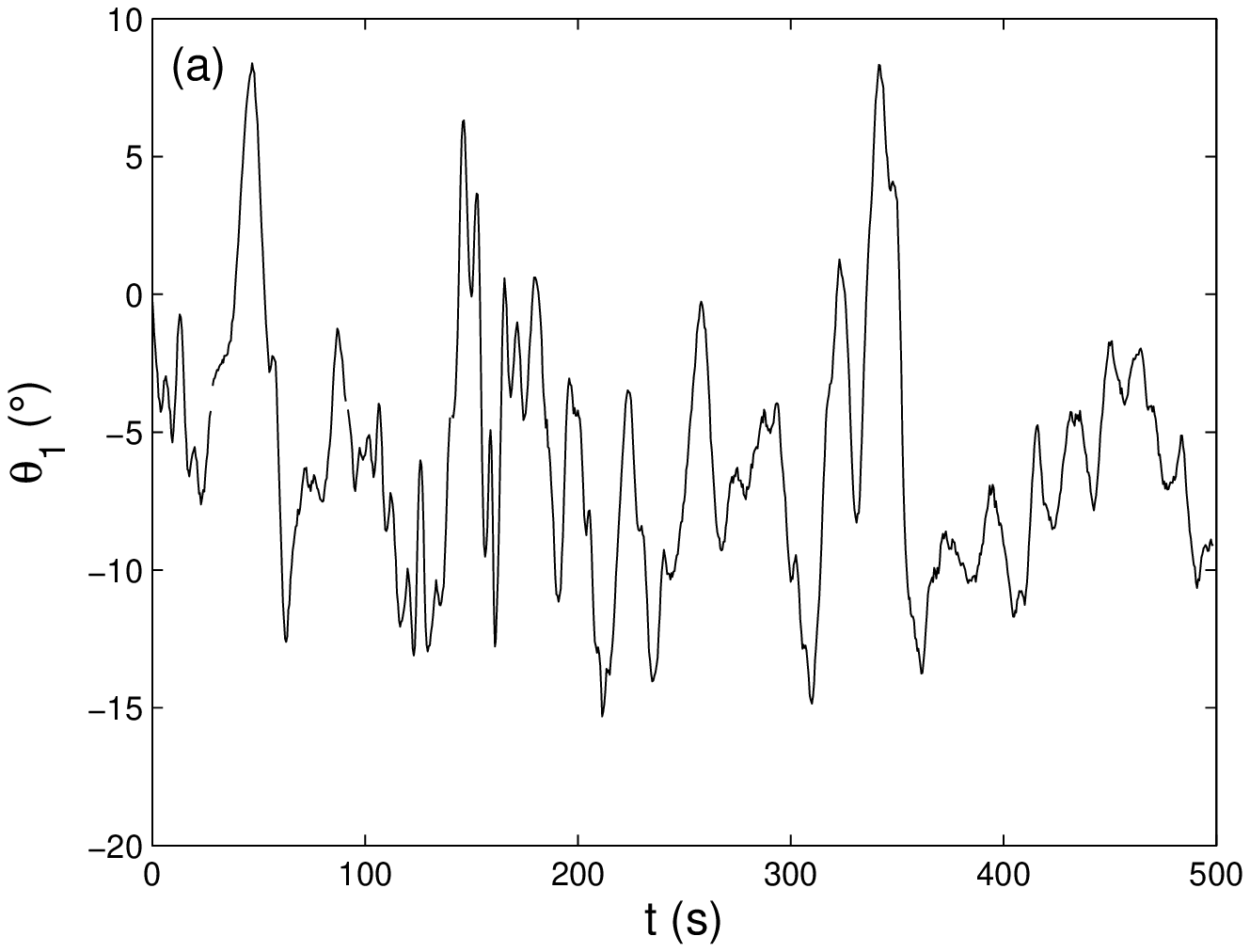}
		\includegraphics[width=0.5\textwidth]{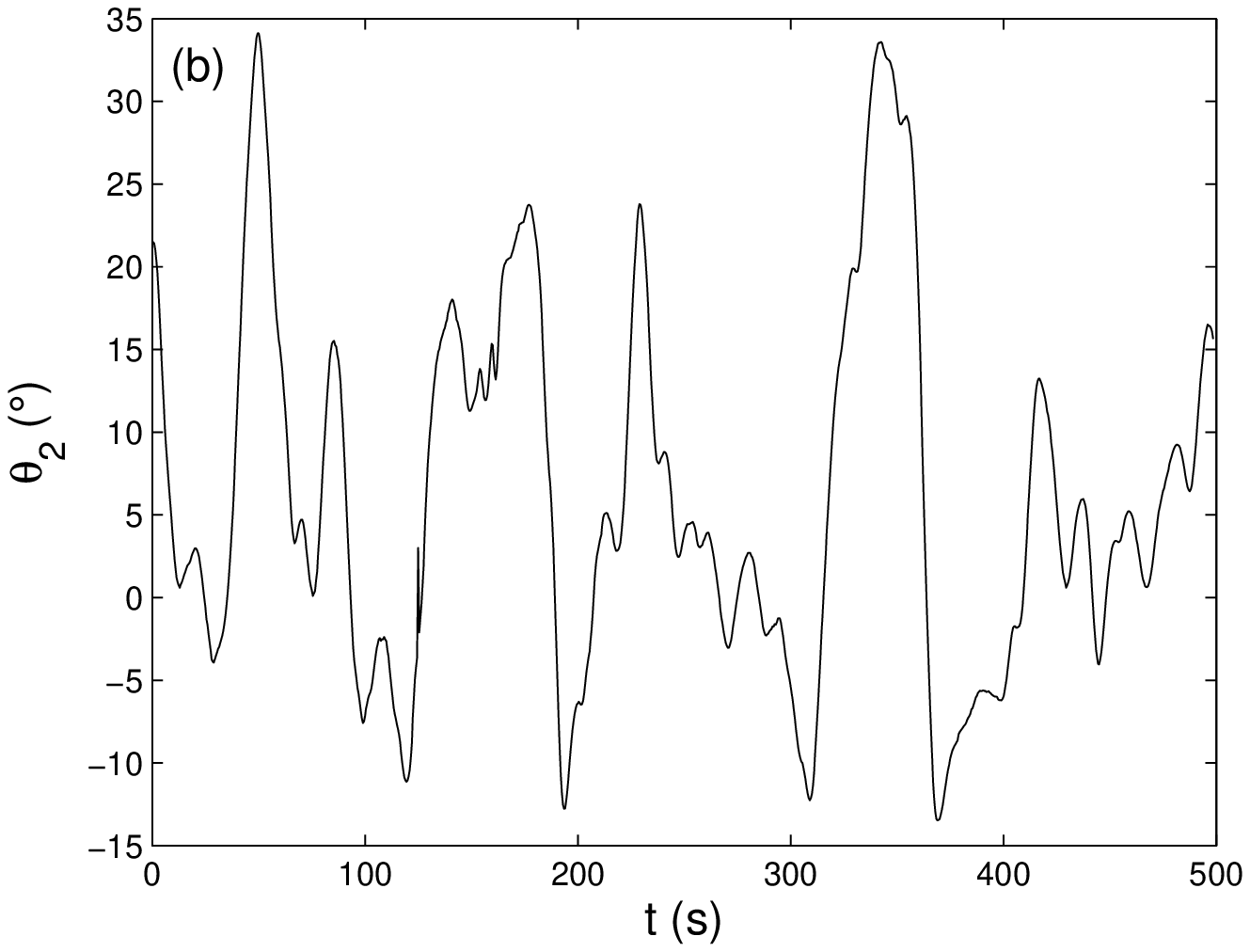}
		\includegraphics[width=0.23\textwidth]{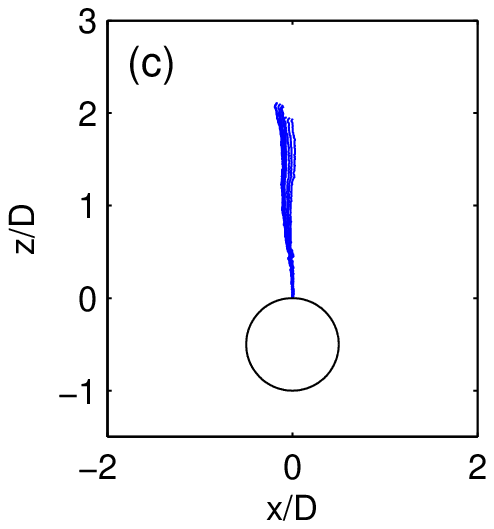}
		\includegraphics[width=0.23\textwidth]{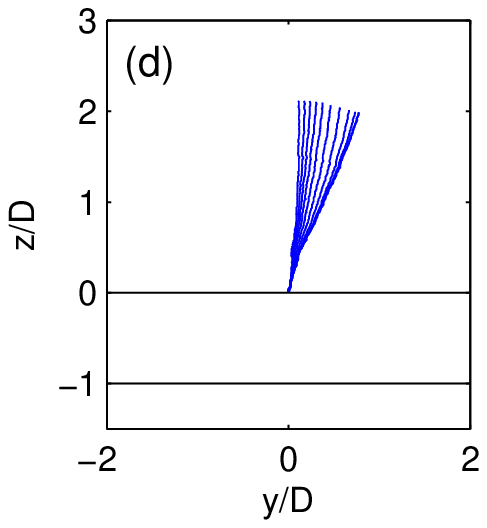}
		\end{center}
\caption{Same as figure \ref{fig:Theta_vs_t_L9.2_Re310} but for $L/D = 2.1$.}
\label{fig:Theta_vs_t_L2.1_Re310}
\end{figure}

In this section, the time evolution of the filament position in the cylinder wake are shown, followed by a study of the mean position of the filament as a function of its length. Figures \ref{fig:Theta_vs_t_L9.2_Re310}, \ref{fig:Theta_vs_t_L4.08_Re310} and \ref{fig:Theta_vs_t_L2.1_Re310} show the time evolution of the angles of deflection, $\theta_1$ and $\theta_2$ and the corresponding filament positions at different times for three filament lengths, namely, $L_s = 9.2$, $L_s = 4.1$ and $L_s = 2.1$ ($Re=310$). 
 For the longest length $L_s=9.2$ shown in figure \ref{fig:Theta_vs_t_L9.2_Re310}, the angles $\theta_1$ and $\theta_2$ oscillate around a value close to $0^\circ$, indicating that the filament is almost vertical (figures \ref{fig:Theta_vs_t_L9.2_Re310}a and \ref{fig:Theta_vs_t_L9.2_Re310}b). This can also be seen in the filament positions presented in figure \ref{fig:Theta_vs_t_L9.2_Re310}(c) for the front view and in figure \ref{fig:Theta_vs_t_L9.2_Re310}(d) for the side view. Figure \ref{fig:Theta_vs_t_L4.08_Re310} shows that for the intermediate filament length, $L_s=4.1$, the angle $\theta_1$ oscillates around a value close to $23^\circ$ and the angle $\theta_2$ oscillates around $80^\circ$  (figures \ref{fig:Theta_vs_t_L4.08_Re310}a and \ref{fig:Theta_vs_t_L4.08_Re310}b). This indicates that the filament is lying down as can be observed in figures \ref{fig:Theta_vs_t_L4.08_Re310}(c) and \ref{fig:Theta_vs_t_L4.08_Re310}(d). Note that for this situation, the detectable part of the filament from the front view is reduced; however, the angle $\theta_2$ clearly shows the collapse of the filament towards the cylinder surface. For the shortest filament $L_s=2.1$, the angles $\theta_1$ and $\theta_2$ are close to $0^\circ$ (figures \ref{fig:Theta_vs_t_L2.1_Re310}a and \ref{fig:Theta_vs_t_L2.1_Re310}b) indicating that the filament is again almost vertical. This can also be observed in the filament position in figure \ref{fig:Theta_vs_t_L2.1_Re310}(c) and \ref{fig:Theta_vs_t_L2.1_Re310}(d).

\begin{figure}
	\begin{center}
		\includegraphics[width=0.5\textwidth]{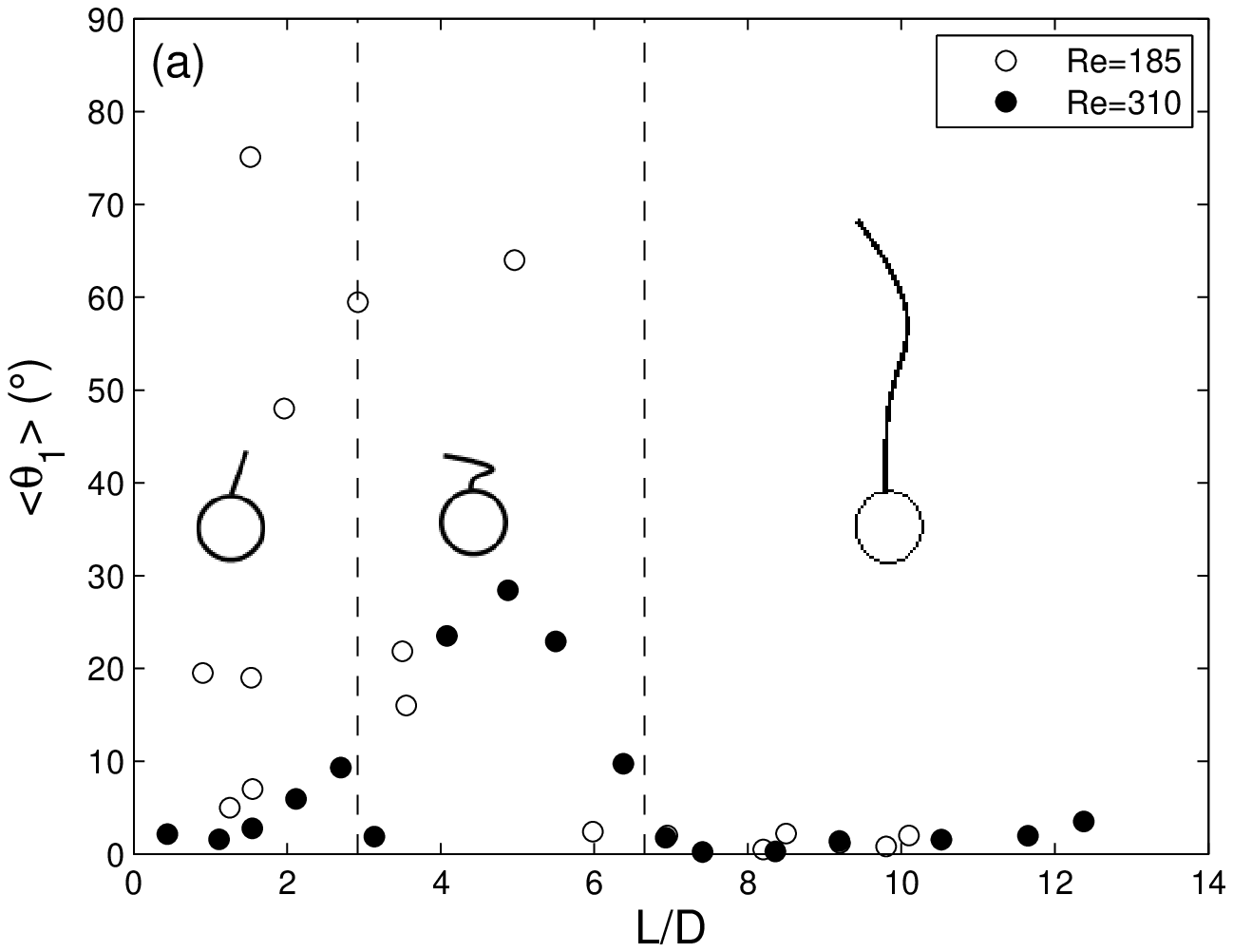}
		\includegraphics[width=0.5\textwidth]{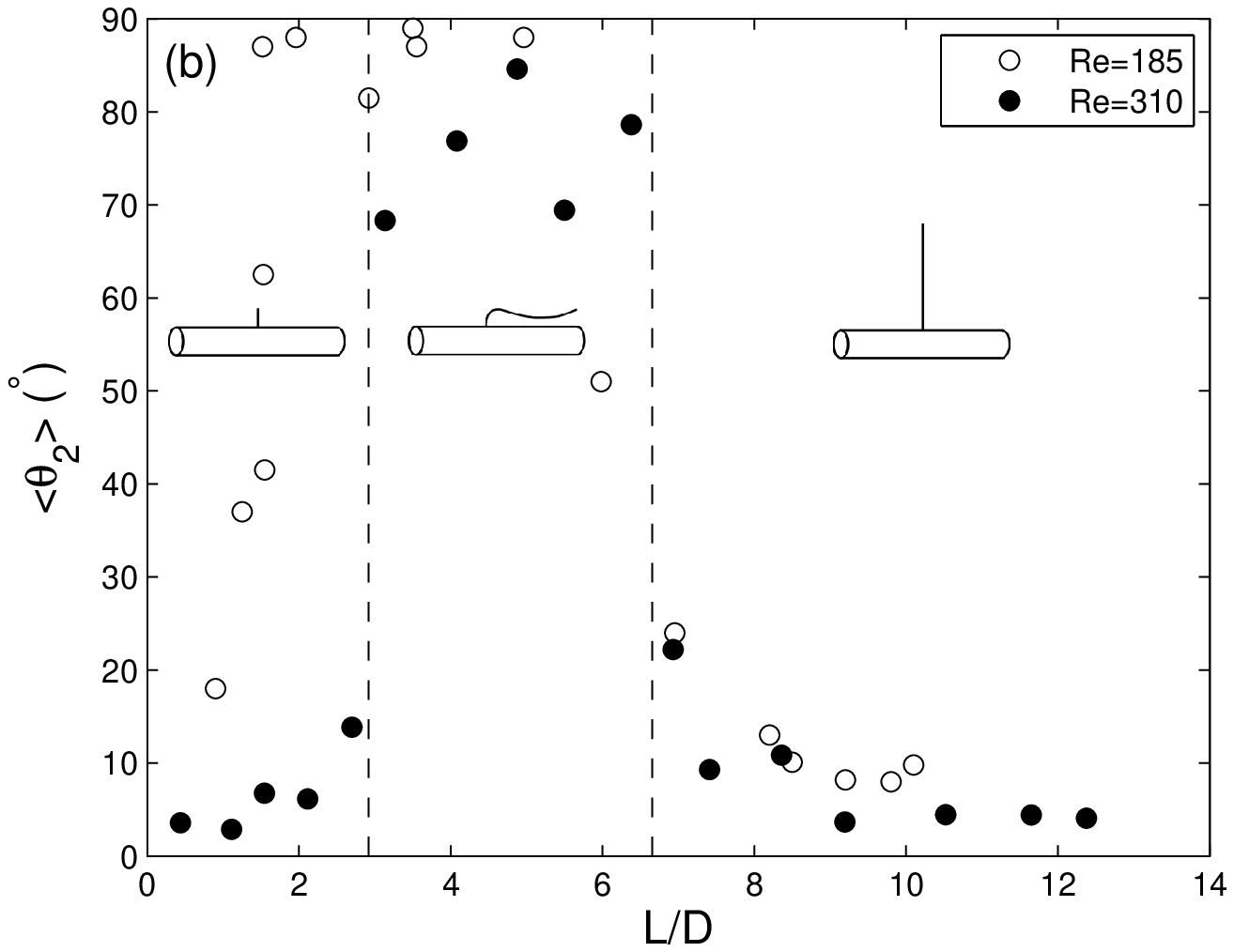}
	\end{center}
\caption{Average angles of deflection (a) $\langle\theta_1\rangle$ and (b) $\langle\theta_2\rangle$ as a function of the filament length for two Reynolds numbers $Re=185$ and $Re=310$.}
\label{fig:Theta_vs_L}
\end{figure}

\begin{figure}
 \begin{center}
			\includegraphics[width=0.5\textwidth]{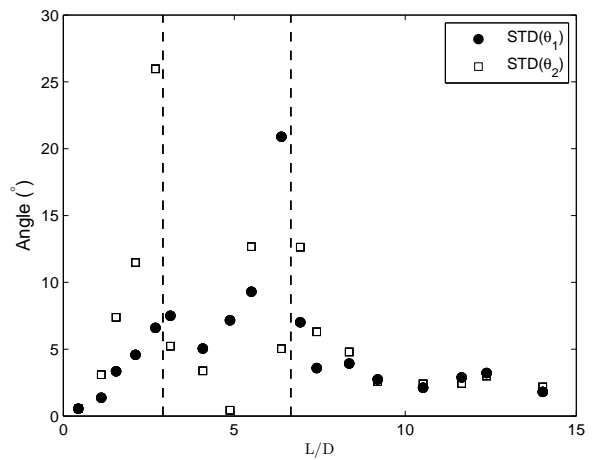}
 \end{center}
	\caption{Standard deviation of the angles $\theta_1$ and $\theta_2$ as a function of the filament length for the $Re=310$ case.}
	\label{fig:RMS_Thetha_vs_L}
 \end{figure}

The average angles of deflection $\langle \theta_1\rangle$ and $\langle\theta_2\rangle$ are shown in figures \ref{fig:Theta_vs_L}(a) and \ref{fig:Theta_vs_L}(b) as a function of $L_s$ for the two Reynolds numbers under investigation.
For filaments longer than $L_s= 6.5$,  $\langle\theta_1\rangle$ and
$\langle\theta_2\rangle$ are smaller than $5^\circ$ and $25^\circ$, respectively.
This indicates that the mean position of the filament is nearly vertical from the front view - i.e. aligned with the incoming flow - and slightly tilted from the side view (schematically shown as inset frames in figure \ref{fig:Theta_vs_L}). Figure \ref{fig:RMS_Thetha_vs_L} presents the standard deviation of the angles $\theta_1$ and $\theta_2$ as a function of the filament length for $Re=310$. For these long filament lengths the standard deviation of the filament oscillations are below $5^\circ$ for both angles and increase close to the bifurcation.

When the filament length is reduced below a critical value, $L_s\sim6.5$, a first bifurcation takes place as the angle $\langle\theta_2\rangle$ increases dramatically to reach values higher than $70^\circ$. The angle $\langle\theta_1\rangle$  also increases at this bifurcation. Thus, in this regime the filament tends to collapse (i.e. fall down) on the cylinder and align with the axis of symmetry of the cylinder (normal to the incoming flow) as schematically shown in figure \ref{fig:Theta_vs_L}.
 Once the filament collapses on one side, it does not flip to the other side. 
 Out of 10 experiments we performed for this particular configuration, the filament  fell down to the left twice and to the right 8 times.  
 In perfect symmetric conditions, one would expect an equal probability for the filament to fall to right and left, however further samples and statistical analysis are required to confirm this.
 For these intermediary filament lengths the standard deviation of the angle $\theta_1$ varies between $5^\circ$ and $21^\circ$; the standard deviation of the angle $\theta_2$ varies between $0.5^\circ$ and $13^\circ$ (figure \ref{fig:RMS_Thetha_vs_L}). The evolution of these amplitudes with the filament length are non-monotonic in this range.

Finally a second bifurcation takes place at $L_s\sim3$ for $Re=310$ and at $L_s\sim1.5$ for $Re=185$; the value of the mean angle $\langle\theta_1\rangle$ is smaller than $20^\circ$(figure \ref{fig:Theta_vs_L}a) and the angle $\langle\theta_2\rangle$ is reduced from values higher than $65^\circ$ to values lower than $20^\circ$ (figure \ref{fig:Theta_vs_L}b). This indicates that the mean position of the filament is still tilted, but tends to realign with the incoming flow as shown schematically in figure \ref{fig:Theta_vs_L}. For these short filament lengths the standard deviation of $\theta_1$ and $\theta_2$ decrease from respectively $7^\circ$ and $25^\circ$ to $0.5^\circ$ for both angles (figure \ref{fig:RMS_Thetha_vs_L}), indicating that the amplitudes of the filament oscillations decrease with the filament length.

\begin{figure}
  \begin{center}
			\includegraphics[width=0.48\textwidth]{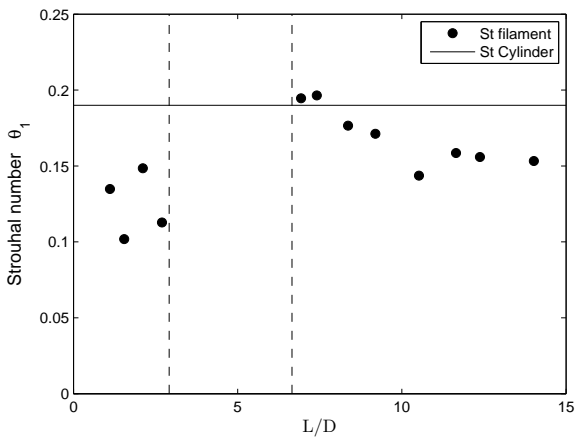}
 \end{center}
	\caption{Strouhal number $St=f_1\frac{d}{U}$ of the oscillation frequency of $\theta_1$ as a function of the filament length for the $Re=310$ case compared to the Strouhal number of the vortex shedding in the cylinder wake at $Re=300$ \citep{Williamson1996}.}
	\label{fig:St_VS_L_Re310}
 \end{figure}

Concerning the frequencies of the filament oscillations, it can be observed from figures \ref{fig:Theta_vs_t_L9.2_Re310}(a), \ref{fig:Theta_vs_t_L9.2_Re310}(b), \ref{fig:Theta_vs_t_L4.08_Re310}(a), \ref{fig:Theta_vs_t_L4.08_Re310}(b),  \ref{fig:Theta_vs_t_L2.1_Re310}(a) and \ref{fig:Theta_vs_t_L2.1_Re310}(b) that several distinct frequencies are contained in each signal. For $\theta_2$ no clear  frequency seem to dominate. For $\theta_1$  on the other hand, the Strouhal number  $St=f_1 D /U$ of the dominant frequency $f_1$ is plotted as a function of the filament length in figure \ref{fig:St_VS_L_Re310} (for $Re=310$). For long filament lengths, $L/D>6.5$, the Strouhal number of $\theta_1$ is  between $0.14$ and $0.2$, which is close to the natural vortex shedding frequency of the cylinder wake $St_{cylinder} \sim 0.19$ (at $Re=300$, see e.g. \cite{Williamson1996}).  Thus, for $L/D>6.5$ the filament oscillation frequency seems to be somewhat synchronized with the vortex shedding in the cylinder wake. Note however that $\theta_1$ contains also other frequencies that are not shown here. For intermediary filament lengths $2< L/D <6.5$ no clear dominant frequency could be identified. Finally, for short filament length $L/D < 2$ the Strouhal number of the filament oscillation frequency is between $0.1$ and $0.15$ and is significantly different from $St_{cylinder}$.  The different dynamical regimes determined from the average angles seem thus to be reflected in the dominant time scale of  $\theta_1$.

We can also note some differences in the bifurcations between the two Reynolds numbers studied here, namely $Re=185$ and $Re=310$. As previously indicated, the main marker of the bifurcations is the angle $\theta_2$ and it can be compared between the two Reynolds numbers on figure \ref{fig:Theta_vs_L}(b). For the higher Reynolds number, $Re=310$, the first bifurcation appears at $L_s\sim 6.5$ and at $L_s\sim 3$ for the second whereas for $Re=185$ the bifurcations appear at a slightly different filament lengths; at $L_s\sim 6$ and $L_s\sim 1.5$. The bifurcations seem also to be more sharp at $Re=310$ compared to  $Re=185$ where it is more continuous. Despite these differences, it is clear that the same bifurcations  are observed for both Reynolds numbers and these bifurcations seem to be robust in this range of Reynolds number.

\section{Discussion}

Although, further quantitative investigations are necessary for confirmation, we believe that it is likely that the bifurcations described above are related to a buckling instability and an IPL instability.
At the Reynolds numbers considered in this study, the
 recirculation region \citep{Williamson1996} can be characterized by its mean length, which decreases from 1.5 to 1 D when the Reynolds number increases from $Re=120$ to $Re=300$ \citep{LimaESilva2003,NishiokaSato1978,ParkEal1998}. Inside this recirculation region we can define a back-flow region where the fluid velocity is opposite to the streamwise direction. This  back-flow region can be modelled as an ellipse  in the $xz$-plane \citep{Lacis2014} and in the three-dimensional case this ellipse is extended in the $z-$direction. The shape of the back-flow region is schematically shown in figure \ref{fig:Schematic_Angles_Theta_1}(a) for the $xz$-plane and in figure \ref{fig:Schematic_Angles_Theta_2}(a) for the $yz$-plane. The behaviour of the filament can be related to this mean flow and may be attributed to two different types of instabilities. 
 

The first instability is a similar mechanism as the two dimensional IPL instability \citep{Lacis2014}, i.e. the pressure force in the back-flow region behind the cylinder, exerts a net normal force $F_n^+$  on the filament, which pushes the filament away from the symmetric vertical position. The part of the filament outside the back-flow-region on the other hand,  experiences a net normal restoring force $F_n^-$, which pushes the filament towards the straight vertical position.  As shown schematically  in figure \ref{fig:Schematic_Angles_Theta_1}(a), the final position of the filament is due to the competition between these two forces, which for sufficiently short filaments results in a tilted mean position with respect to the direction of the incoming flow.
The key difference to the work of \citet{Lacis2014} is that in the present three-dimensional case, we have observed that the filament falls down on the cylinder in the vertical $yz$-plane containing the cylinder axis (in contrast to the $xz$-plane). In the $yz$-plane, the back-flow region has on average a rectangular form (in contrast to the elliptic shape in the $xz$-plane). As a consequence, the restoring normal force $F_n^-$ outside the back-flow region becomes increasingly small as  $\theta_2$ increases and once the straight position of the filament is unstable no other force can prevent the filament to lie down on the cylinder.  Although, physically an IPL-type of instability could be responsible for the first bifurcation, the critical value of  filament length near $L_s=6.5$ is much larger than the critical values ($L_s\sim 2$) reported in previous work on the IPL instability and also much larger than the length of the recirculation region  which is $\sim 1$ D at these Reynolds numbers \citep{LimaESilva2003,NishiokaSato1978,ParkEal1998}.

The second instability that could induce the collapse of the filament onto the cylinder surface in the $xz$-plane is a buckling-type of mechanism. This instability arises out of the competition between the tangential forces $F_t^+$ and $F_t^-$ (shown schematically  in figure \ref{fig:Schematic_Angles_Theta_2}a) on the filament inside and outside the back-flow region. As a consequence of these  forces, part of the filament outside the back-flow region is stretched by the flow and the part of the filament inside the back-flow region is compressed.  If the filament is sufficiently short, the compressive force will dominate, which in turn may result in a buckling instability of the filament. For the second bifurcation at $L_s\sim 2$, where the filament tends to realign with the flow, the filament is unlikely to undergo a buckling instability as the critical load for the onset of the buckling instability scales as $1/L^2$.

The  above discussion suggests that the first bifurcation at $L_s=6.5$ could be dominated by a buckling-type of instability. 
Further studies, in particular by estimating the loads acting on the filament, are necessary to confirm this statement and to quantitatively assess the mechanisms underlying the bifurcations.

\section{Conclusions}

This experimental study has focused on the behavior of a filament attached to
the rear of a three dimensional circular cylinder placed in a uniform incoming flow. The length of
the filament was varied and for each length the mean position of the filament was
determined. For lengths, $L_s$ longer than $6.5$ the mean position of the filament
is nearly aligned with the incoming flow. For intermediate
lengths, $2 < L_s < 6.5$, it was found that the filament lies down on the cylinder,
i.e. the mean position of the filament is aligned with the axis of the
cylinder.  Finally for $L/D$ shorter than $2$ the mean position is tilted in both
directions with respect to the incoming flow. To the best of our knowledge these two bifurcations of a flexible filament in wake flows have not been observed nor reported in earlier work.

 \bibliographystyle{spbasic}      

\bibliography{Biblio}

\end{document}